\documentclass[ aip,jmp,preprint]{revtex4-1}
\usepackage{amsmath}
\usepackage{graphicx}
\usepackage{dcolumn}
\usepackage{bm}
\usepackage{subfigure}

\begin{document}

\title[Localized modes in mini-gaps ]{Localized modes in mini-gaps opened by
periodically modulated intersite coupling in two-dimensional nonlinear
lattices }
\author{Goran Gligori\'{c}}
\email{goran79@vinca.rs}
\author{Aleksandra Maluckov}
\author{Ljup\v co Had\v{z}ievski}
\author{Boris A. Malomed}
\date{\today}

\begin{abstract}
Spatially periodic modulation of the intersite coupling in two-dimensional
(2D) nonlinear lattices modifies the eigenvalue spectrum by opening \textit{%
mini-gaps} in it. This work aims to build stable localized modes in the new
bandgaps. Numerical analysis shows that single-peak and composite two- and
four-peak discrete static solitons and breathers emerge as such modes in
certain parameter areas inside the mini-gaps of the 2D \textit{superlattice}
induced by the periodic modulation of the intersite coupling along both
directions.The single-peak solitons and four-peak discrete solitons are
stable in a part of their existence domain, while unstable stationary states
(in particular, two-soliton complexes) may readily transform into robust
localized breathers.
\end{abstract}

\maketitle

\affiliation{P$^{*}$ group, Vin\v ca Institute of Nuclear Sciences,
University of Belgrade, P. O. B. 522,11001 Belgrade, Serbia}

\affiliation{Department of Physical Electronics, School of
Electrical Engineering, Faculty of Engineering, Tel Aviv University,
Tel Aviv 69978, Israel}

\preprint{AIP/123-QED}




\begin{quotation}
Nonlinear lattices offer a possibility to create a vast variety of
self-trapped (i.e., spontaneously localized) discrete wave packets, alias
discrete solitons. They are supported by the stable balance between the
onsite nonlinearity and discrete diffraction in the lattice. In optics,
discrete solitons have been observed in one- and two-dimensional (1D and 2D)
arrays of nonlinear waveguides. In recent years, such \textit{photonic
lattices} have been implemented as permanent structures, or optically
induced as virtual ones, using various materials, including those with
cubic, quadratic, photorefractive, and liquid-crystal nonlinearities. In
these systems, discrete solitary modes (fundamental and multi-peak solitons,
2D vortex solitons, etc.) are observed under both the self-focusing
(in-phase states) and defocusing nonlinearity. In the former case, the
discrete solitons have an in-phase structure (in particular, the fundamental
single-peak solitons are represented by real positive solutions), while in
the latter case solitons exist in the \textit{staggered} form, with
alternating signs of the discrete field at adjacent sites. Discrete solitons
are also known in many other fields, such as Bose-Einstein condensates
(BEC), electric transmission lines, solid-state lattice media, polymer
molecules, etc.

In this work, we address the formation and ensuing dynamics of localized
structures in 2D photonic \textit{superlattices}, induced by a relatively
long-wave periodic modulation of the intersite coupling imposed on the
underlying lattice with the onsite cubic nonlinearity. In addition to
photonics, similar BEC-trapping settings can be built in the form of
modulated optical lattices, by means of a superposition of laser beams
illuminating the condensate. We demonstrate that the periodic modulation of
the intersite coupling opens narrow gaps in the linear spectrum of the 2D
superlattices, and thus enables the creation of new solitary modes (\textit{%
gap solitons}) in these \textit{mini-gaps}. Some of these modes are robust
(not only static ones, but also periodically oscillating \textit{breathers}%
), therefore they can be experimentally created in the photonic and
matter-wave (BEC) settings.
\end{quotation}

\section{Introduction}

Discrete solitons represent self-trapped states in nonlinear lattice
systems. They result from the interplay between the lattice diffraction and
material nonlinearity. In optics, these states have been experimentally
observed in both one- and two-dimensional (1D and 2D) nonlinear waveguiding
arrays. Such photonic lattices have been built as permanent structures, or
induced as virtual ones, in a variety of optical media, including those with
cubic, quadratic, photorefractive (saturable), and liquid-crystal
nonlinearities \cite{Silb, Silb2, Lederer, Segev, christo, Ki1D, Ki2D,
vortex, Chen, Kip, review, Jena, Jena2}. Similar spatially periodic trapping
structures for quasi-discrete states of matter waves, based on optical
lattices (induced by the interference of laser beams illuminating the
Bose-Einstein condensate, BEC), have been created too \cite{Inguscio,261,
263, 262, Zoller, Sengstock}. In addition, discrete solitons are known in
chains of micromechanical oscillators \cite{micro}, liquid crystals \cite%
{andrea}, biological macromolecules \cite{biol,Scott}, and in other settings.

Discrete fundamental \cite{Segev,christo} and vortical \cite{Ki2D, vortex,
Chen,Denz} solitons in 2D nonlinear waveguide arrays were first observed in
biased photorefractive crystals. Properties of fundamental discrete solitons
were studied in detail theoretically too \cite{PGK, ICFO, Kevr2D, mi2D}. In
these systems, solitons exist in unstaggered (i.e., in-phase) and staggered
(i.e., with phase shift $\pi $ between adjacent lattice sites) forms, under
the self-focusing (SF) and self-defocusing (SDF)\ onsite nonlinearity,
respectively. Discrete solitary vortices were also studied in detail by
means of numerical methods \cite{kevrvortex1,Dimitri,johanson2d,PGK}. In
particular, attention has been drawn to the 2D band structure associated
with photonics lattices and respective aspects of the stability of 2D
discrete solitons, see, e.g., Ref. \cite{134}. Specific features of lattice
solitons in BEC models were addressed in Refs. \cite{154, 2D-BEC, nashvortex}%
.

Discrete localized modes were also studied in inhomogeneous 1D lattices,
subject to a quasiperiodic spatial modulation of the intersite coupling
constant \cite{Sukhoru}, as well as with an inhomogeneous onsite
nonlinearity \cite{nashexp}. The modulation of the lattice coupling constant
merely implies a varying spacing between the sites, as the coupling constant
depends on it exponentially. An interesting finding is that the strength of
the onsite SDF nonlinearity, which grows from the center to periphery of the
1D lattice at any rate faster than the distance, $|n|$, supports solitons of
the unstaggered type, which are impossible in the uniform lattice with the
SDF sign of the nonlinearity \cite{nashexp}. Equivalently, solitons of the
staggered type are supported by the growing onsite SF nonlinearity, which is
not possible either in the uniform lattice. It was demonstrated too that
unstaggered solitons exist in the lattice with homogeneous onsite SDF
nonlinearity, if the coupling constant decays fast enough at $|n|\rightarrow
\infty $ \cite{nashexp}.

In this paper, we aim to extent the study of discrete solitons to 2D
inhomogeneous lattices, namely, to those with periodic modulation of the
intersite coupling. These settings may be considered as \textit{superlattices%
}, which, in the general case, are defined as structured created by imposing
relatively long-wave periodic spatial modulations onto the underlying
lattice \cite{super}. The model, based on the corresponding discrete
nonlinear Schr\"{o}dinger (DNLS) equations with the SF onsite nonlinearity,
is introduced in Section II. First, we study linear properties of the
modulated 2D lattices, and demonstrate opening of mini-gaps in the
corresponding linear spectrum, where self-trapping of new types of discrete
solitons may be expected. For the sake of the completeness, in subsection
III.A we briefly present results of the study of self-trapped modes
(fundamental and vortex solitons) residing in the semi-infinite spectral gap
in the superlattice, and compare them to the corresponding results in
uniform lattices. The main topic of this work, which is presented in
subsection III.B, is the creation of stable discrete solitons in the
mini-gaps. We find families of fundamental solitons in mini-gaps, as well as
two- and four-soliton complexes. While the fundamental solitons are stable,
numerical simulations demonstrate that the complexes evolve into localized
breathing structures. The paper is concluded by Section IV.

\section{The model}

\subsection{Basic equations}

\begin{figure}[h]
\centering {\includegraphics[width=9cm]{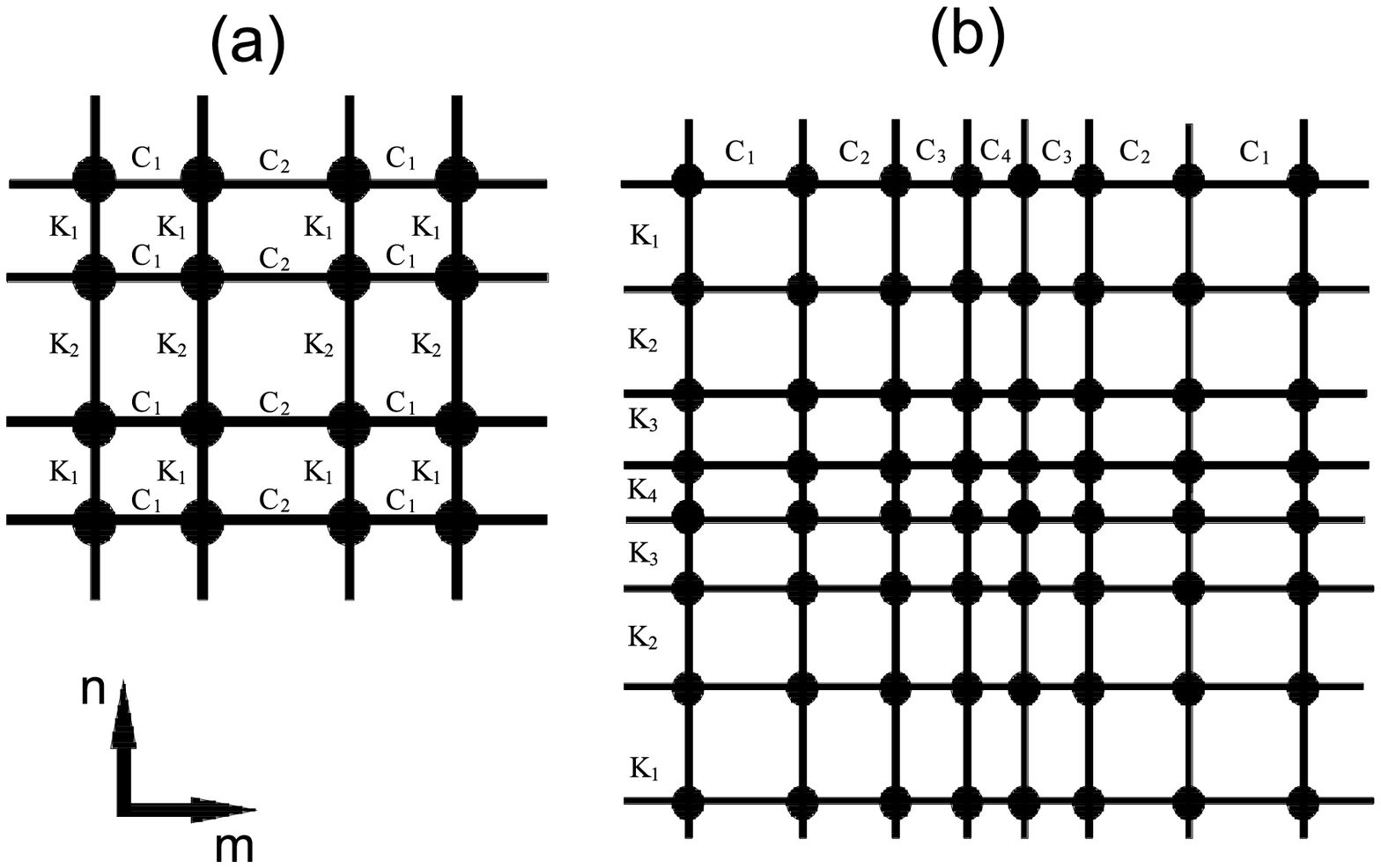}} \caption{(a)
The schematic plot of the lattice with corresponding to modulation
pattern (\protect\ref{eq1a}) with $Q_{a}=Q_{b}=\protect\pi $, see
also Eq. (\protect\ref{eq2a}). (b) A fragment of the superlattice
corresponding to $Q_{a}=Q_{b}=\protect\pi /3$. Different values of
the coupling constants are designated in the plots.} \label{fig1}
\end{figure}

The 2D discrete model is based on the following DNLS equation, with the
cubic onsite SF nonlinearity, for complex field amplitudes $\psi _{m,n}$:
\begin{equation}
i\frac{d\psi _{m,n}}{dz}+C_{m,n}\left( \psi _{m+1,n}+\psi _{m-1,n}\right)
+K_{m,n}\left( \psi _{m,n+1}+\psi _{m,n-1}\right) +\gamma \left\vert \psi
_{m,n}\right\vert ^{2}\psi _{m,n}=0,  \label{eq1}
\end{equation}%
where the horizontal and vertical coupling constants are modulated,
respectively, along the horizontal and vertical directions, as shown in Fig. %
\ref{fig1}:
\begin{eqnarray}
C_{m,n} &=&C_{0}\left[ 1+\Delta _{1}\cos (Q_{a}m)\right] \equiv C_{m}  \notag
\\
K_{m,n} &=&C_{0}\left[ 1+\Delta _{2}\cos (Q_{b}n)\right] \equiv K_{n}.
\label{eq1a}
\end{eqnarray}%
This setting can be easily implemented in photonic lattices by properly
selecting distances between the constituent waveguides, as well as for the
BEC loaded into deep optical lattices, shaped as shown in Fig. \ref{fig1}.
In the former case, evolution variable $z$ is the propagation distance along
individual waveguides, while in the matter-wave (BEC) realization, $z$ is
replaced by time $t$.

In addition, we also considered the 2D lattice with another modulation
pattern, corresponding to the horizontal and vertical couplings periodically
modulated along the vertical and horizontal directions, respectively:
\begin{eqnarray}
C_{m,n} &=&C_{0}\left[ 1+\Delta _{1}\cos (Q_{a}n)\right]  \notag \\
K_{m,n} &=&C_{0}\left[ 1+\Delta _{2}\cos (Q_{b}m)\right] .  \label{eq1b}
\end{eqnarray}%
Unlike the model based on Eq. (\ref{eq1}), the analysis has not revealed any
dynamically stable localized structures in Eq. (\ref{eq1b}), except for the
usual onsite solitons in the semi-infinite spectral gap, and it is not
straightforward to implement the latter model experimentally. Therefore, we
here focus on Eq. (\ref{eq1a}).

The results are presented below for $\gamma =+1$ (which corresponds to the
SF nonlinearity), $C_{0}=1$, and $\Delta _{1}=\Delta _{2}=0.5$ in Eqs. (\ref%
{eq1}) and (\ref{eq1a}). Furthermore, $Q_{a}=Q_{b}=\pi /3$ is fixed in Eq. (%
\ref{eq1a}), which corresponds to the superlattice displayed in Fig. \ref%
{fig1}(b), unless it is specified otherwise. Comparison with many results
obtained at other values o the parameters, including those with $Q_{a}\neq
Q_{b}$, demonstrates that this case adequately represents the generic
situation.

Stationary solutions to Eqs. (\ref{eq1}) with real propagation constant $%
-\mu $ (or chemical potential for the BEC) are looked for as
\begin{equation}
\psi _{m,n}(z)=e^{-i\mu z}U_{m,n},  \label{uU}
\end{equation}%
where stationary discrete function $U_{m,n}$ obeys the following equation:
\begin{equation}
\mu U_{m,n}+C_{m}\left( U_{m+1,n}+U_{m-1,n}\right) +K_{n}\left(
U_{m,n+1}+U_{m,n-1}\right) +|U_{m,n}|^{2}U_{m,n}=0,  \label{eq2}
\end{equation}%
with the coupling constants defined as per Eq. (\ref{eq1a}). The
power of the
discrete soliton is defined as usual,%
\begin{equation}
P=\sum_{m,n}|U_{m,n}|^{2}.  \label{P}
\end{equation}

Stationary equation (\ref{eq2})\ was solved by means of a numerical
algorithm based on the modified Powell minimization method \cite{mi2D}.
Stability of the so found discrete solitons was checked, in the framework of
the linear stability analysis, by numerically solving the corresponding
eigenvalue equation for modes of small perturbations. Finally, the evolution
equation (\ref{eq1}) was directly simulated by dint of the Runge-Kutta
procedure of the sixth order, cf. Ref. \cite{mi2D}. The simulations were
used to verify the stability properties predicted by the linear analysis.

\subsection{The linear spectrum}

The linearized version of Eq. (\ref{eq2}) is
\begin{equation}
\mu U_{m,n}+C_{m}\left( U_{m+1,n}+U_{m-1,n}\right) +K_{n}\left(
U_{m,n+1}+U_{m,n-1}\right) =0.  \label{eq3}
\end{equation}%

Its eigenvalue (EV) spectrum can be derived analytically for \textit{binary
lattices}, which are characterized by alternating values of the coupling
constants ($Q_{a}=Q_{b}=\pi $), see Fig. \ref{fig1}(a):
\begin{eqnarray}
C_{m} &=&C_{0}\left[ 1+\left( -1\right) ^{m}\Delta _{1}\right] =1\pm
0.5\equiv C_{1(2)},  \notag \\
K_{m} &=&C_{0}\left[ 1+\left( -1\right) ^{n}\Delta _{2}\right] =1\pm
0.5\equiv K_{1(2)},  \label{eq2a}
\end{eqnarray}%
with the top and bottom signs corresponding, severally, to even and odd
values of $m$ or $n$. The corresponding solution for amplitudes in this case
can be looked for as
\begin{equation}
\left( a_{m,n}\,,b_{m,n},c_{m,n},\,d_{m,n}\right) =\left( A,B,C,D\right)
\exp \left[ {i(\kappa _{a}m+\kappa _{b}n)}\right] ,  \label{ABCD}
\end{equation}%
where $\kappa _{a},~\kappa _{b}$ are the Bloch wavenumbers in the
$m$ and $n$ directions, while $a,b,c,d$ pertain to four sets of
sites distinguished by the four different values of the coupling
constants in Eq. (\ref{eq2a}). The substitution of ansatz
(\ref{ABCD}) into Eq. (\ref{eq3}) leads to the following
eigenvalue equation:
\begin{gather}
16C_{1}^{2}C_{2}^{2}\,\cos ^{4}\kappa _{a}-4\,\left( \cos ^{2}\kappa
_{a}\right) \left[ (C_{1}^{2}+C_{2}^{2})\mu ^{2}+8C_{1}C_{2}K_{1}K_{2}\,\cos
^{2}\kappa _{b}\right]   \notag \\
+\left( \mu ^{2}-4K_{1}^{2}\,\cos ^{2}\kappa _{b}\right) (\mu
^{2}-4K_{2}^{2}\,\cos ^{2}\kappa _{b})=0.  \label{EV}
\end{gather}%
For $\kappa _{a,b}=\pi /2$, Eq. (\ref{EV}) degenerates to $\mu ^{4}=0$, at
which point the gaps get closed. In this situation, we have found a few
different families of nonstationary localized solutions, which radiate due
to coupling to linear lattice modes in the absence of the gap.

In the general case, the eigenvalue problem was solved
numerically. It has been found that, in addition to the
semi-infinite gap, the spectrum of the superlattice contains new
narrow (\textit{mini}-) gaps, in which the linear Bloch waves do
not exist, while their nonlinear counterparts are modulationally
unstable, thus opening the way to create localized structures
(discrete gap solitons) with the propagation constant falling into
the mini-gaps \cite{Salerno}, via the interplay of the
nonlinearity, discreteness and modulated intersite coupling. Our
main objective here is to demonstrate that some of such gap
solitons are stable. They can be observed experimentally
\cite{gapexp}, and may be used to control the light propagation in
photonic structures. It is relevant to mention that mini-gaps, and
gap solitons existing in them, are known in continual nonlinear
models of Bragg
supergratings, i.e., gratings subject to a long-wave modulation \cite%
{superBG}.

\begin{figure}[h]
\center\includegraphics [width=13cm]{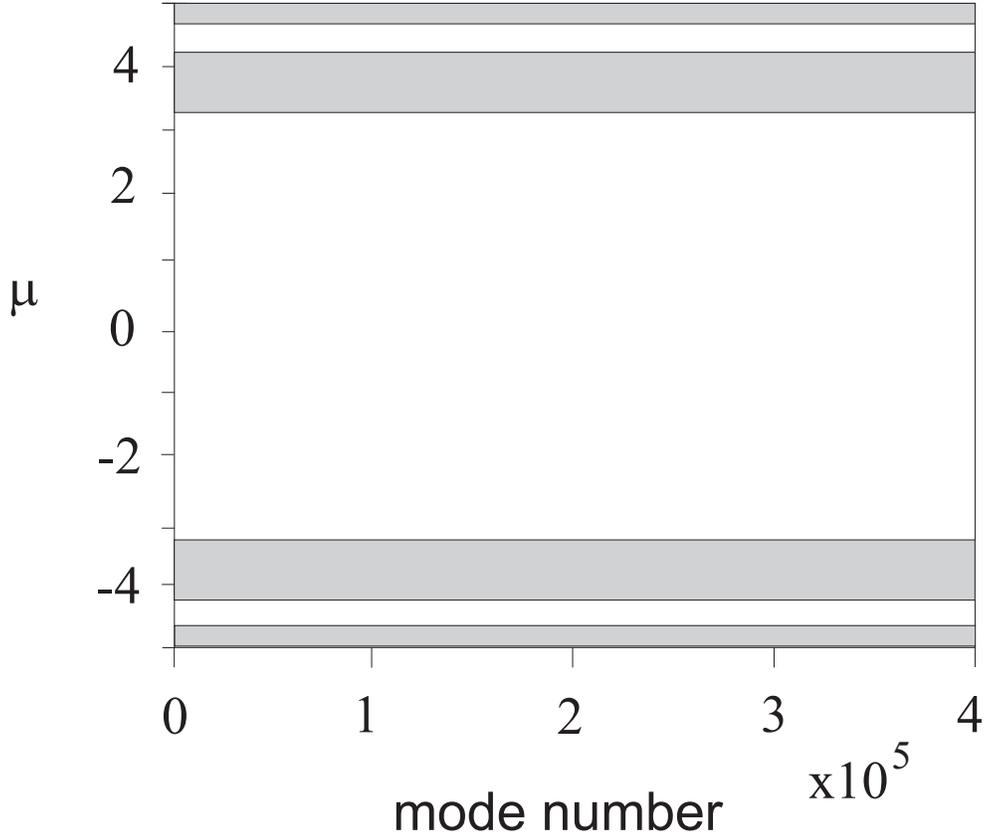} \caption{The linear
spectrum for the lattice with $Q_{a}=Q_{b}=\protect\pi /3 $ and
$\Delta _{1}=\Delta _{2}=0.5$. The spectrum features
semi-infinite gaps at (approximately) $|\protect\mu |>4.73$, and mini-gaps around $|%
\protect\mu |=4$ (both presented by gray areas). Similar spectra,
with slightly different positions and widths of the gaps, are
obtained for other values of $Q_{a},Q_{b}$. } \label{fig2}
\end{figure}

As an illustration, the linear spectrum for the lattice with $%
Q_{a}=Q_{b}=\pi /3$ is presented in Fig. \ref{fig2}. Clearly visible are
mini-gaps around $|\mu |=4$, which are placed symmetrically with respect to $%
\mu =0$. Extensive numerical calculations, performed for lattices
with different ratios between $Q_{a}$ and $Q_{b}$, produce similar
spectra.

\section{Localized modes}

It is known that 2D nonlinear lattices with the uniform intersite coupling
give rise to fundamental bright discrete solitons of the unstaggered type,
and different types of vortices in the parameter region corresponding to the
semi-infinite gap in the corresponding linear spectrum \cite{PGK, ICFO,
Kevr2D, mi2D, kevrvortex1, Dimitri, johanson2d}. In the case of the 2D
lattices with the SDF onsite nonlinearity, bright solitons are produced by
the staggering transformation \cite{PGK}.

In comparison with the uniform lattice, the periodic modulation of the
intersite coupling opens the mini-gaps, in which new discrete solitons are
expected, as said above. We demonstrate below that discrete solitons
residing in the semi-infinite gap of the modulated lattices are not
significantly altered by the spatially periodic inhomogeneity, while
completely novel species of staggered discrete solitons are found in the
mini-gaps.

\subsection{Soliton families in the semi-infinite gap}

\begin{figure}[h]
\center\includegraphics [width=13cm]{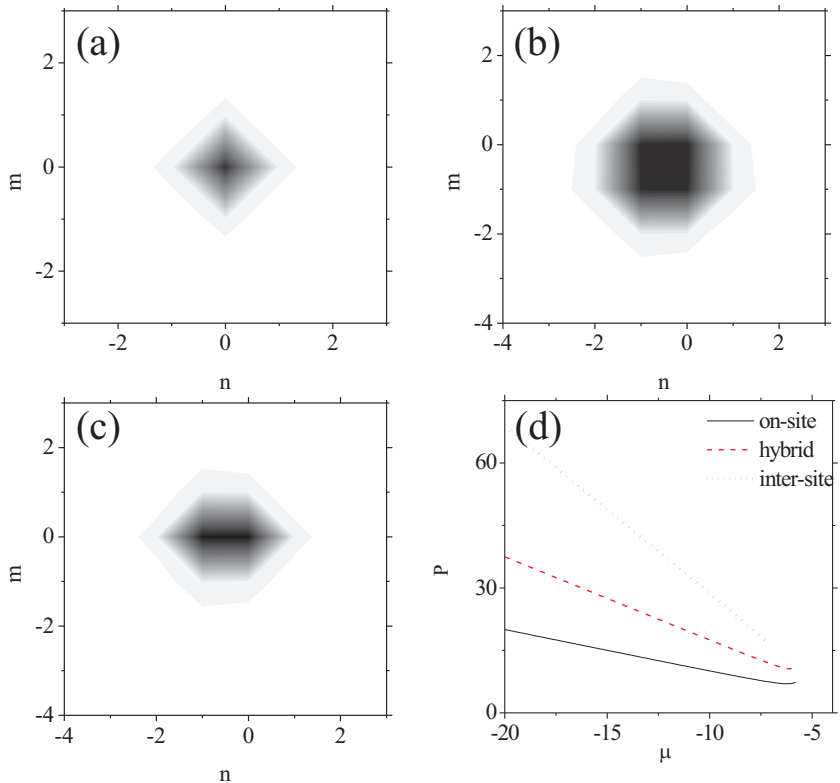}
\caption{(Color online) Amplitude profiles of the 2D solitons belonging to
the semi-infinite gap in the superlattice created by periodic modulation (%
\protect\ref{eq1b}) with $Q_{a}=Q_{b}=\protect\pi /3$: (a) onsite-centered,
(b) inter-site-centered, and (c) hybrid fundamental solitons. Plot (d) shows
the dependence of the soliton's norm $P$ vs. $\protect\mu $ for the
fundamental modes (solid line - onsite, dashed line - hybrid, dotted line -
inter-site); the norm is defined as per Eq. (\protect\ref{P}).}
\label{fig3}
\end{figure}

\begin{figure}[h]
\center\includegraphics [width=8cm]{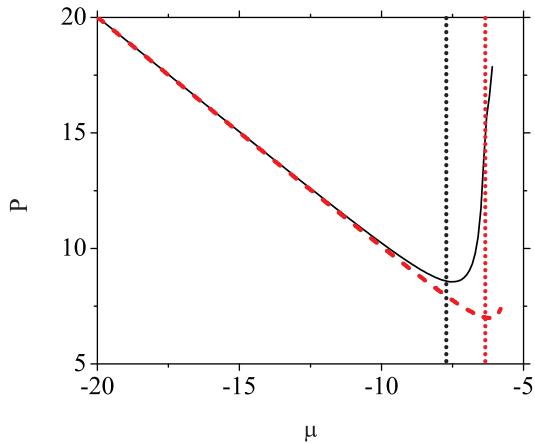}
\caption{(Color online) The $P(\protect\mu )$ dependence for 2D onsite
solitons belonging to the semi-infinite gap: black solid and red dashed
lines correspond to the uniform and periodically modulated ($Q_{a}=Q_{b}=%
\protect\pi /3$) lattices, respectively. The stability region for the
uniform and modulated lattices are located, severally, on the left of the
dotted black and red vertical lines.}
\label{unifpor}
\end{figure}

Three types of fundamental unstaggered solitons have been found in
the semi-infinite gap ($\mu <-4.73$ in Fig. \ref{fig2}): onsite,
hybrid, and intersite ones, see Fig. \ref{fig3}. They feature
dynamical properties similar to those of their counterparts in the
2D uniform lattice. In particular, solely the onsite family is
stable, in almost the entire existence region, see Fig.
\ref{unifpor}. The modulation of the lattice only slightly extends
the area where the stable onsite modes are found.

Vortex solitons with topological charge $S=1$ and $S=2$ are formed too in
the semi-infinite gap. In terms of their stability and dynamics, they are
also similar to their counterparts in the 2D uniform lattice. In particular,
vortices with topological charges $S=1$ and $S=2$ \cite%
{kevrvortex1,Dimitri,PGK} are stable in certain parts of their existence
region. The anisotropic intersite coupling in the modulated 2D lattice
affects the shape of the solitons, as illustrated in Fig. \ref{fig3} for the
intersite and hybrid fundamental solitons.

\subsection{Soliton families in the mini-gaps}

\subsubsection{Fundamental solitons}

\begin{figure}[h]
\center\includegraphics [width=13cm]{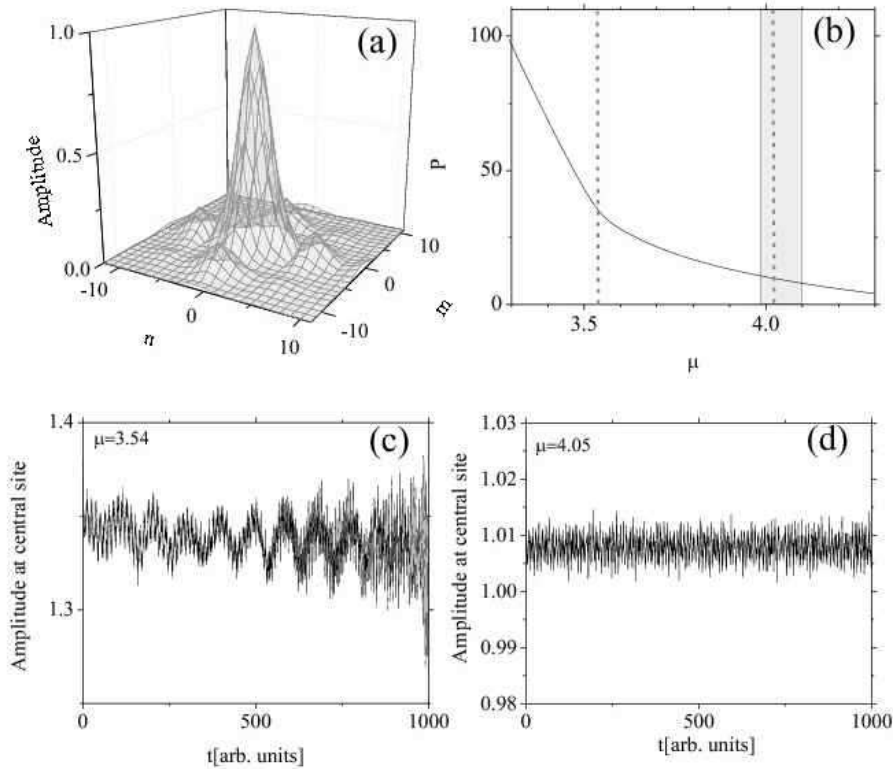}
\caption{(Color online) Solitons in the mini-gap opened around $\protect\mu %
=4$, cf. Fig. 3. (a) A stable soliton found for $\protect\mu =4.05$ [this
value of $\protect\mu $ is designated by the right green vertical dashed
line in panel (b)]. (b) The solitons's norm $P$ vs. $\protect\mu $. The gray
rectangle designates the stability region of the soliton's type presented in
plot (a). Plots (c) and (d) show the amplitude at the central site vs. time,
produced by direct simulations of Eq. (\protect\ref{eq1}) at $\protect\mu %
=3.54$ [a strongly unstable soliton, designated by the left green vertical
dashed line in panel (b)], and $\protect\mu =4.05$ [the stable soliton
displayed in (a)].}
\label{fig4}
\end{figure}

Several different single-soliton families have been found in the mini-gaps.
We here consider the single one, which is stable in a part of its existence
region, see Fig. \ref{fig4}(a), while all the other families are completely
unstable. The $P(\mu )$ dependence for the soliton family is shown in Fig. %
\ref{fig4}(b).  By approaching both edges of the mini-gap the
single soliton families disappear in the sense that corresponding
localized patterns cannot be created. In the lower bound $P$
vanishes, while in the upper grows. In both cases background is
characterized by highly irregular amplitudes with corresponding
magnitude (small and high, respectively) without clearly
distinguishing localized structure.  It is found that, except for
the light-gray area, $3.98<\mu <4.1$ [the rectangle in Fig.
\ref{fig4} (b)], where the calculation of the EVs for small
perturbations shows that the solitons are stable, in other domains
they are unstable. These predictions are corroborated by direct
simulations, as shown in \ref{fig4}(c,d). In particular, the
strong instability observed in Fig. \ref{fig4}(c) is accounted for
by purely real EVs.

\subsubsection{Solitons complexes}

\begin{figure}[h]
\center\includegraphics [width=14cm]{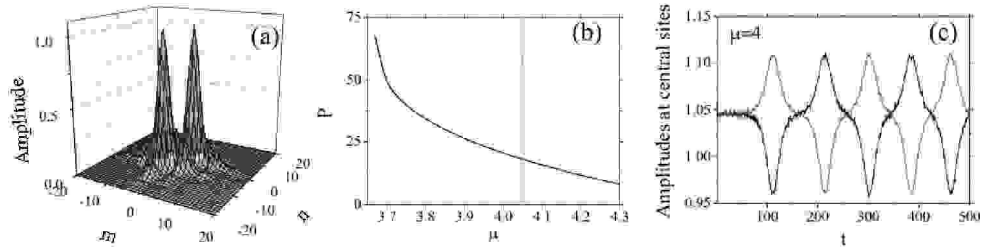}
\caption{(Color online) Dynamics of two-soliton complexes in the mini-gap
opened around $\protect\mu =4$. (a) The profile of the two-peak complex ($%
\protect\mu =4$). (b) The norm of the complex, $P$, vs. $\protect\mu $ (c)
The amplitudes of the constituent solitons vs. time, in the complex
designated by the vertical line in (b).}
\label{fig5}
\end{figure}

In addition to dynamically stable single-soliton gap modes, we have found
different bound states of solitons in the minigap. First, two-soliton
complexes are built of two identical single-soliton constituents, see Fig. %
\ref{fig5}. It is found that the two-soliton complex keeps its compactness
and the staggered structure in the course of perturbed evolution, which
gives rise to regular amplitude oscillations, as seen in \ref{fig5}(c). In
direct simulations, the two-soliton complex is actually found to be more
robust than predicted by the linear-stability analysis, which demonstrate
the presence of pure real EVs for small perturbations, i.e., weak
instability of such complexes.

\begin{figure}[h]
\center\includegraphics [width=14cm]{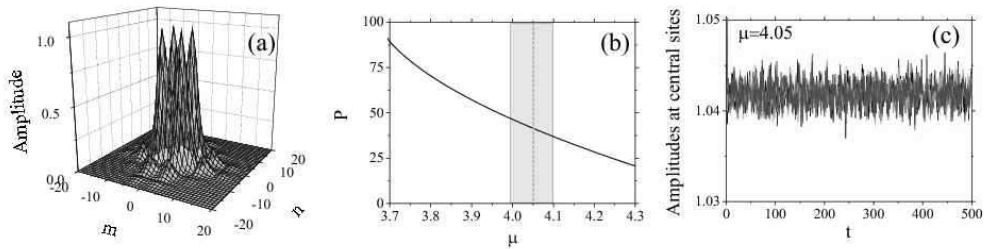}
\caption{(Color online) Four-soliton complexes in the mini-gap. (a) The
profile of the stable complex at $\protect\mu =4.05$. (b) The norm, $P$, vs.
$\protect\mu $, for the family of the four-soliton complexes. The gray
rectangle denotes the stability region for solitons of the type presented in
panel (a). (c) Amplitudes of the constituent solitons vs. time, for the
stable complex designated by the vertical line in (b), in the course of its
perturbed evolution.}
\label{fig6}
\end{figure}

On the other hand, four-soliton complexes, which are built of four identical
in-phase individual solitons (with zero phase shifts between them), see Fig. %
\ref{fig6}, feature stability properties similar to those of the
constituent solitons, as shown in Fig. \ref{fig6}(b). In
particular, the instability of the four-soliton complex is
accounted for by purely real EVs, when they are present, and the
complex is stable in the absence of such eigenvalues, which is
corroborated by direct simulations. Thus, the in-phase
four-soliton complexes are essentially more stable than their
two-soliton counterparts. On the other hand, out-of-phase
composites were always found to be unstable, independently on the
number of constituent solitons. Note that both multi-soliton
families cannot be formed in the neighborhood of the mini-gap
edges. The corresponding soliton branches are changed by solution
branches corresponding to more or less irregular background
similarly to the case of fundamental solitons.

In the case of the SDF nonlinearity, single-soliton modes and their two- and
four-soliton bound states can be generated by the staggering transformation
from their unstaggered counterparts found for the SF nonlinearity in the
mini-gap opened around $\mu =-4$, see Fig. \ref{fig2}. Naturally, these
staggered modes feature the same stability and dynamics as their
counterparts which were considered above.

\subsection{Other solutions}

Direct simulations clearly demonstrate that all the modes observed
in the mini-gaps are immobile in the 2D lattice: in direct
simulations, the application of a kick to stable modes, in any
direction with respect to the underlying superlattice, causes
oscillations of the kicked solitons, but not their progressive
motion (not shown here in detail). In fact, such simulations,
although they fail to produce motile discrete solitons, corroborate
their stability against positional perturbations.

The lack of the mobility of the discrete solitons in the present model is
not surprising, taking into regard the general property of immobility of 2D
discrete solitons supported by the cubic nonlinearity \cite{Hadi}, which is,
in turn, explained by the vulnerability of the continuum-limit counterparts
of such solitons to the collapse. As a result, a soliton will be compressed
by the (quasi-) collapse until it will become strongly localized, on few
lattice sites, i.e., strongly pinned to the lattice, which provides for its
stabilization but prevents it from being mobile.

It is known that the mobility of discrete solitons may be enhanced by
special modes of \textquotedblleft management", i.e., by the application of
time-periodic [or $z$-periodic, in terms of Eq. (\ref{eq1})] modulation of
the local nonlinearity (at least, in 1D settings) \cite{Cuevas}. The
consideration of such mechanisms may be a subject for a separate work, being
beyond the scope of the present one.

Lastly, in the superlattice with the spatially periodic modulation of the
intersite coupling defined as per Eq. (\ref{eq1b}), new gaps in the linear
spectrum open too. In those gaps, different stationary localized solutions
can be found. The linear-stability analysis predicts that all these states
are subject to an oscillatory instability, while in direct simulations some
of them may seem as robust breathers, which is explained by the saturation
of the instability with the increase of the oscillation amplitude.

\section{Conclusion}

We have demonstrated that periodic modulation of the intersite coupling
opens narrow gaps in the linear spectrum of the 2D square-shaped lattices,
offering the possibility to create new species of discrete solitons in these
mini-gaps, if the onsite cubic nonlinearity acts in the system. A part of
the family of fundamental solitons and the respective four-soliton complexes
are dynamically stable. Some other modes, which are unstable, develop into
robust localized breathers. These modes can be experimentally created in
arrays of nonlinear optical waveguides and in BEC trapped in a deep optical
lattice, be means of currently available techniques  \cite{Segev}-\cite{Chen}%
. It may be interesting to extend the analysis to 2D lattices with different
geometries, such as triangular, honeycomb, and quasi-periodic.

\acknowledgments G.G., A.M., and Lj.H. acknowledge support from the Ministry
of Education, Science and Technological Development of the Republic of
Serbia (Project III45010).


\begin{thebibliography}{99}
\bibitem{Silb} H. S. Eisenberg, Y. Silberberg, R. Morandotti, A. R. Boyd,
and J. S. Aitchison, Phys. Rev. Lett. \textbf{81}, 3383(1998).

\bibitem{Silb2} A. A. Sukhorukov, Y. S. Kivshar, H. S. Eisenberg, and Y.
Silberberg, IEEE J. Quantum Electron. \textbf{39}, 31 (2003).

\bibitem{Lederer} D. N. Christodoulides, F. Lederer, and Y. Silberberg,
Nature \textbf{424}, 817 (2003).

\bibitem{Segev} J. W. Fleischer, M. Segev, N. K. Efremidis, and D. N.
Christodoulides, Nature \textbf{422}, 147 (2003).

\bibitem{christo} J. W. Fleischer, T. Carmon, M. Segev, N. K. Efremidis, and
D. N. Christodoulides, Phys. Rev. Lett. \textbf{90}, 023902 (2003).

\bibitem{Ki1D} D. Neshev, E. Ostrovskaya, Y. Kivshar, and W. Kr\'{o}%
likowski, Opt. Lett. \textbf{28}, 710 (2003).

\bibitem{Ki2D} D. N. Neshev, T. J. Alexander, E. A. Ostrovskaya, Y. S.
Kivshar, H. Martin, I. Makasyuk, and Z. Chen, Phys. Rev. Lett. \textbf{92},
123903 (2004).

\bibitem{vortex} J. W. Fleischer, G. Bartal, O. Cohen, O. Manela, M. Segev,
J. Hudock, and D. N. Christodoulides, Phys. Rev. Lett. \textbf{92}, 123904
(2004).

\bibitem{Denz} R. Fischer, D. Trager, D. N. Neshev, A. A. Sukhorukov, W. Kr%
\'{o}likowski, C. Denz, and Y. S. Kivshar, Phys. Rev. Lett. \textbf{96},
023905 (2006); B.Terhalle, T. Richter, A. S. Desyatnikov, D. N. Neshev, W. Kr%
\'{o}likowski, F. Kaiser, C. Denz, and Y. S. Kivshar, \textit{ibid}. \textbf{%
101}, 013903 (2008).

\bibitem{Chen} D. Song, C. Lou, L. Tang, X. Wang, W. Li, X. Chen, K. J. H.
Law, H. Susanto, P. G. Kevrekidis, J. Xu, and Z. Chen, Opt. Exp. \textbf{16}%
, 10110 (2008).

\bibitem{Kip} J. Hukriede, D. Runde, and D. Kip, J. Phys. D: Appl. Phys.
\textbf{36}, R1 (2003).

\bibitem{review} F. Lederer, G. I. Stegeman, D. N. Christodoulides, G.
Assanto, M. Segev, and Y. Silberberg, Phys. Rep. \textbf{463}, 1 (2008).

\bibitem{Jena} M. Heinrich, R. Keil, F. Dreisow, A. T\"{u}nnermann, A.
Szameit, and S. Nolte, Appl. Phys. B \textbf{104}, 469 (2011).

\bibitem{Jena2} U. R\"{o}pke, H. Bartelt, S. Unger, K. Schuster, and J.
Kobelke, Appl. Phys. B \textbf{104}, 481 (2011).

\bibitem{Inguscio} F. S. Cataliotti, L. Fallani, F. Ferlaino, C. Fort, P.
Maddaloni, and M. Inguscio, New J. Phys. \textbf{5}, 71 (2003).

\bibitem{261} B. Eiermann, Th. Anker, M. Albiez, M. Taglieber, P. Treutlein,
K.-P. Marzlin, M. K. Oberthaler, Phys. Rev. Lett. \textbf{92}, 230401 (2004).

\bibitem{263} T. Anker, M. Albiez, B. Eiermann, M. Taglieber, M. Oberthaler,
Opt. Exp. \textbf{12}, 11 (2004).

\bibitem{262} Th. Anker, M. Albiez, R. Gati, S. Hunsmann, B. Eiermann, A.
Trombettoni, M. K. Oberthaler, Phys. Rev. Lett. \textbf{94}, 020403 (2005).

\bibitem{Zoller} D. Jaksch and P. Zoller, Ann. Phys. \textbf{315}, 52 (2005).

\bibitem{Sengstock} P. Windpassinger and K. Sengstock, Rep. Prog. Phys.
\textbf{76}, 086401 (2013).

\bibitem{micro} M. Sato, B. E. Hubbard, and A. J. Sievers, Rev. Mod. Phys.
\textbf{78}, 137 (2006).

\bibitem{andrea} A. Fratalocchi, G. Assanto, K. A. Brzdakiewicz, and M. A.
Karpierz, Opt. Exp. \textbf{13}, 1808 (2005).

\bibitem{biol} A. S. Davydov and N. I. Kislukha, Phys. Stat. Solid. B
\textbf{59}, 465 (1973).

\bibitem{Scott} A. C. Scott, Phys. Rev. A \textbf{26}, 578 (1982).

\bibitem{PGK} P. G. Kevrekidis, "The Discrete Nonlinear Schr\"{o}dinger
Equation: Mathematical Analysis, Numerical Computations, and Physical
Perspectives" (Springer: Berlin and Heidelberg, 2009).

\bibitem{ICFO} Y. V. Kartashov, V. A. Vysloukh, and L. Torner, Prog. Opt.
\textbf{52}, 63 (2009).

\bibitem{Kevr2D} P. G. Kevrekidis, K. O. Rasmussen, and A. R. Bishop, Phys.
Rev. E \textbf{61}, 2006 (2000); P. G. Kevrekidis, H. Susanto, and Z. Chen,
Phys. Rev. E \textbf{74}, 066606 (2006)

\bibitem{mi2D} G. Gligori\'c, A. Maluckov, M. Stepi\'c, Lj. Had\v zievski,
and B. A. Malomed, Phys. Rev. A \textbf{81}, 013633 (2010).

\bibitem{kevrvortex1} B. A. Malomed, and P. G. Kevrekidis, Phys. Rev. E
\textbf{64}, 026601 (2001); P. G. Kevrekidis, B. A. Malomed, Z. Chen, and D.
J. Frantzeskakis, Phys. Rev. E \textbf{70}, 056612 (2004).

\bibitem{Dimitri} P. G. Kevrekidis, B. A. Malomed, Z. Chen, and D. J.
Frantzeskakis, Phys. Rev. E \textbf{70}, 056612 (2004).

\bibitem{johanson2d} M. \"{O}ster and M. Johansson, Phys. Rev. E \textbf{73}%
, 066608 (2006).

\bibitem{134} N. K. Efremidis, J. Hudock, D. N. Christodoulides, J. W.
Fleischer, O. Cohen, and M. Segev, Phys. Rev. Lett. \textbf{91}, 213906
(2003).

\bibitem{154} N. K. Efremidis and D. N. Christodoulides, Phys. Rev. A
\textbf{67}, 063608 (2003).

\bibitem{2D-BEC} A. Gubeskys and B. A. Malomed, Phys. Rev. A \textbf{76},
043623 (2007).

\bibitem{nashvortex} G. Gligori\'{c}, A. Maluckov, M. Stepi\'{c}, Lj. Had%
\v{z}ievski, and B. A. Malomed, J. Phys. B: At. Mol. Opt. Phys. \textbf{43},
055303 (2010).

\bibitem{Sukhoru} A. A. Sukhorukov, Phys. Rev. Lett. \textbf{96}, 113902
(2006).

\bibitem{nashexp} G. Gligoric, A. Maluckov, Lj. Had\v{z}ievski, and B. A.
Malomed, Phys. Rev. E \textbf{88}, 032905 (2013).

\bibitem{super} C. P. Collier, T. Vossmeyer, and J. R. Heath, Ann. Rev.
Phys. Chem. \textbf{49}, 371 (1998); A. Wacker, Phys. Rep. \textbf{357}, 1
(2002); A. A. Lebedev, Semicond. Sci. Tech. \textbf{21}, R17 (2006).

\bibitem{Salerno} V. V. Konotop and M. Salerno, Phys. Rev. \textbf{65},
021602 (2002).

\bibitem{gapexp} B. Eiermann, Th. Anker, M. Albiez, M. Taglieber, P. Treutlein, K.-P.
Marzlin, and M. K. Oberthaler, Phys. Rev. Lett. \textbf{92},
230401 (2004).





\bibitem{superBG} P. J. Y. Louis, E. A. Ostrovskaya, and Y. S. Kivshar,
Phys. Rev. A \textbf{71}, 023612 (2005); M. A. Porter and P. G. Kevrekidis,
SIAM J. Appl. Dyn. Syst. \textbf{4}, 783 (2005); M. Porter, P. G.
Kevrekidis, R. Carretero-Gonz\'{a}lez, and D. J. Frantzeskakis, Phys. Lett.
A \textbf{352}, 210 (2006); K. Yagasaki, I. M. Merhasin, B. A. Malomed, T.
Wagenknecht, and A. R. Champneys, Europhys. Lett. \textbf{74}, 1006 (2006).

\bibitem{Hadi} H. Susanto, P. G. Kevrekidis, R. Carretero-Gonz\'{a}lez, B.
A. Malomed, and D. J. Frantzeskakis, Phys. Rev. Lett. \textbf{99}, 214103
(2007).

\bibitem{Cuevas} J. Cuevas, B. A. Malomed, and P. G. Kevrekidis, Phys. Rev.
E \textbf{71}, 066614 (2005).
\end{thebibliography}
\end{document}